\documentclass[conference]{IEEEtran}
\usepackage[utf8]{inputenc}
\usepackage[usenames, dvipsnames]{color}
 
\ifCLASSINFOpdf
  \usepackage[pdftex]{graphicx}
 \DeclareGraphicsExtensions{.pdf,.jpeg,.png}
\else
 \usepackage[dvips]{graphicx}
 \DeclareGraphicsExtensions{.eps}
\fi
\usepackage{amsmath}
\usepackage{algorithmic}
\usepackage{array}
\usepackage{tabu}
\ifCLASSOPTIONcompsoc
 \usepackage[caption=false,font=normalsize,labelfont=sf,textfont=sf]{subfig}
\else
\usepackage[caption=false,font=footnotesize]{subfig}
\fi
\graphicspath{{./figures/}}
\usepackage{flushend}
\usepackage{url}
\IEEEoverridecommandlockouts
\begin{document}
\title{Evaluating Marijuana-Related Tweets On Twitter$^1$}



\author{\IEEEauthorblockN{Anh Nguyen$^2 $\IEEEauthorrefmark{2},
Quang Hoang\IEEEauthorrefmark{1},
Hung Nguyen\IEEEauthorrefmark{1},
Dong Nguyen$^3$\IEEEauthorrefmark{1}, and  
Tuan Tran\IEEEauthorrefmark{3} 
 }
\IEEEauthorblockA{\IEEEauthorrefmark{1}Saolasoft Inc., Centennial, CO 880211, USA\\
Email: {\{qhoang, hnguyen, dnguyen\}@saolasoft.com}}


\IEEEauthorblockA{\IEEEauthorrefmark{2}University of Denver, Denver, CO 80208, USA\\
Email: anh.n.nguyen@du.edu}
\IEEEauthorblockA{\IEEEauthorrefmark{3}Sullivan University, Louisville, KY 40205,  USA\\
Email: ttran@sullivan.edu}
\vspace{-0.3in}
\thanks{$^1$This research will be presented in IEEE-CCWC 2017}
\thanks{$^2$This research was done during the author's intern at Saolasoft Inc.}
\thanks{$^3$Corresponding author}
}
\maketitle

\begin{abstract}
This paper studies marijuana-related tweets in social network Twitter. We collected more than 300,000 marijuana related tweets during November 2016 in our study. Our text-mining based algorithms and data analysis unveil some interesting patterns including: (i) users' attitudes (e.g., positive or negative) can be characterized by the existence of outer links in a tweet; (ii) 67\% users use their mobile phones to post their messages while many users publish their messages using third-party automatic posting services; and (3) the number of tweets during weekends is much higher than during weekdays. Our data also showed the impact of the political events such as the U.S. presidential election or state marijuana legalization votes on the marijuana-related tweeting frequencies.

\end{abstract}

\IEEEpeerreviewmaketitle
\section{Introduction}

 As a microblogging website allowing users to post messages of 140 characters or less called ``tweets'', Twitter has become the biggest daily source of news, public opinions, and personal discussions. For example, on the day of the 2016 U.S. presidential election, Twitter had nearly 40 million messages sent by midnight that day \cite{nytimes_url}. Twitter has 310 million monthly active users, posting hundreds of millions of tweets per day \cite{twitter_url}. People on Twitter share, exchange, and discuss any events of their life including various issues relating to their health conditions or health-related behaviors. This phenomenal leads to a research area among public health scientists to achieve public health analytic outcomes by harnessing the vast amount of publicly available health-related data.

That said, there is a growing interest in exploring the wide range of topics over social media Twitter for epidemiology and surveillance. The study of Culotta et al. \cite{culotta2013lightweight} uses the Twitter corpus to estimate influenza rates and the sale volume of alcohol with high accuracy. The researchers in \cite {myslin2013using} reveals a content and sentiment analysis of tobacco-related Twitter messages and categorizes tobacco-relevant posts with the focus on emerging products like hookah and electronic cigarettes among users. Xu et al. \cite{xu2016leveraging} observe the usage patterns of the terms ``cancer", ``breast cancer", ``prostate cancer", and ``lung cancer" between Caucasian and African American groups on Twitter to understand their knowledge and awareness about specific topics in real-time. Apparently, 
the social media-based surveillance highlight the potential that online user-generated data can be a very powerful and valuable medium for monitoring and tracking the health-related issues and health risk behaviors towards addictive substances like alcohol or tobacco of the public community in short time. 

Marijuana has been legalized to be used for the medical and recreational purpose in 8 states Colorado, Washington, Alaska, Oregon, California, Nevada, Maine, and Massachusetts, and Washington, D.C. Despite some medicinal benefits, it is an indisputable fact that marijuana usage has exerted a myriad of detrimental impacts on the public health. For example, according to data provided by the U.S. National Survey on Drug User and Health \cite{macleod2004psychological}, youth with poor academic outcomes were more than four times as likely to have consumed marijuana in the past year than youth with an average of higher grades. Additionally, the study in \cite{fried2002current} also reveals a strong correlation between marijuana usage and the decline in intelligence quotient (IQ). Cannabis possession and consumption are still illegal under the federal law because of its significant health and safety risks to people, especially young individuals \cite{url_mmj_wh}. Thus, the threats that the marijuana usage poses to the public health and our society should be taken into consideration seriously. Consequently, surveillance of actual marijuana use and concerns would be necessary and useful for legislators to impose appropriate public health laws.

Our research leverages Twitter's public data of marijuana-related tweets exchanged among Twitter users to reveal hidden patterns of marijuana related aspects. Particularly, we collected more than 300,000 marijuana-related tweets during November 2016 in our study. We use the unigram and bigram of marijuana related hashtags to compute the word frequencies. Furthermore, we apply some text-mining sentiment techniques to analyze the users' attitudes based on their tweets. Our data indicates a strong correlation between tweets with outer links and positive attitudes, and between the number of marijuana tweets with political events in November 2016. 



\begin{figure}
\centering 
\includegraphics[width=1.0\linewidth]{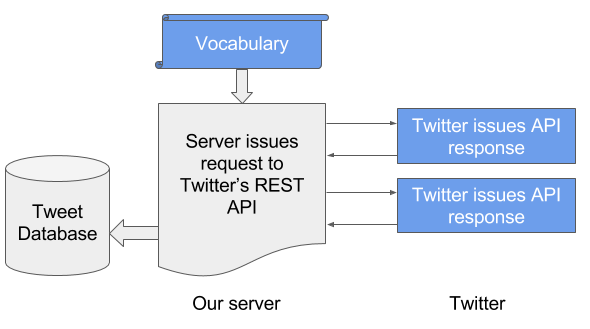} 
\caption{The workflow of tweet collecting server which aggregates the tweets according to the vocabulary of marijuana-related keywords. The server is written in Python.}
\label{fig:twitter_streaming} 
\end{figure}

\section{Related Work}
Together with the rapid expansion of social media including Facebook, Instagram, or Snapchat, microblogging websites such as Twitter have evolved to become an enormous source of various kind of information. Many studies have considered utilizing these sites for event monitoring purposes and event-based surveillance system. For instance, using garnered information from Twitter, the authors in \cite{pak2010twitter} build a sentiment classifier based on n-gram model, that is able to determine positive, negative and neutral emotions of Twitter users. With the same purpose, \cite{agarwal2011sentiment} conducts their research with three types of models: Unigram model, feature-based model, and tree kernel-based model. Besides, Wang et al. \cite{wang2014hurricane} using Twitter corpus unravel a high correlation of some hashtags and tweets posted by users living in the locations affected by the 2012 Hurricane Sandy with its movement. Researchers have also begun to investigate various ways of automatically collecting training Twitter data. Consider \cite{kouloumpis2011twitter} as an example, they take advantages of Twitter hashtags as the training data to train their three-way sentiment classifiers.

Studies on Twitter data are also noticeable in health care sector. For example, the work of \cite{li2016discovering}, for the first time, seeks to recommend relevant Twitter hashtags for health-related keywords based on distributed language representations. Analyzing data from Twitter, blogs, and forums, the authors in \cite{denecke2013exploit} make an attempt to detect hints to public health threats as well as monitor population's health status. In 2012, the researchers in \cite{paul2012model} introduced a new approach to discovering a large number of meaningful ailment description by using machine learning and natural language processing techniques. In \cite{cavazos2015twitter}, the authors make an effort to examine the sentiment and themes of marijuana-related chatters on Twitter sent by influential Twitter users and to describe the demographics of these Twitter users. Another worth noting work is by \cite{cavazos2014characterizing} in which the authors estimate user demographic characteristics based the content of tweets of a popular pro-marijuana Twitter handle and its followers. However, the studies do not consider other meaningful properties of tweets' metadata, such as geographical features, external links, user's device types, etc., and their correlation with each other and with other social phenomena.

\section{Data Collection and Dataset}

\subsection{Data Collection}
We collected and processed more than 300,000 marijuana-related tweets in the English language, posted during November 2016. First of all, we built the marijuana vocabularies, i.e., the list of marijuana-related keywords, with the help of Online Slang Dictionary (\textit{http://onlineslangdictionary.com}). Next, we developed a data collection tool that garnered all tweets containing one or more marijuana-related terms from cities and states in the United States. The workflow of the data collection mechanism is illustrated in Fig. \ref{fig:twitter_streaming}. The server, written in Python, interacts with the Twitter Search API. While the Twitter Search API usually serves Tweets from the past week, our system allows us to bypass this limitation of time constraints by basically replicating the way the Twitter Search engine works on browsers. The server calls the API by command: \textit{``https://twitter.com/i/search/timeline?f=realtime"} with following parameters: 
\begin{itemize}
\item \textit{q}: a query text in which searched tweets will contain. It is a word and phrase relating to marijuana, pre-stored in our vocabulary.
\item \textit{since}: the lower bound of the posting date of searched tweets.
\item \textit{until}: the upper bound of the posting date of searched tweets.
\item \textit{lang}: the language of searched tweets.
\end{itemize}
This API retrieves a list of matched tweets in the form of an HTML string. The server then extracts useful data from HTML string and save them to our tweet database server. 

\subsection{Data Description and Processing}
The resulting tweets are stored in a NoSQL tweet database which includes $316,191$ documents relating to marijuana. Each document represents a tweet with 15 different fields extracted from the JSON object resulted from our data collection process, including username, URL, external links, text, the number of retweets and favorites, keyword, state, posted time, types of devices, etc. Also, those records are linked into a different table of N-gram sequences processed from tweet text. We use unigrams (1-gram) and bigrams (2-gram) for generating text mining clouds. In term of sentiment analysis, we utilized machine learning techniques from a third party to code the Tweets for different types: positive sentiment about marijuana, negative sentiment about marijuana and neutral sentiment about marijuana. In addition, since hashtags (symbol \#) are likely to be used before prime keywords or phrases in a tweet, we extract hashtags from Tweets as well. We also count the total number of external links of each user and analyze the users who post most tweets which contain external links.  

\begin{figure*}
\centering 
\subfloat[]{\includegraphics[width=.9\columnwidth]{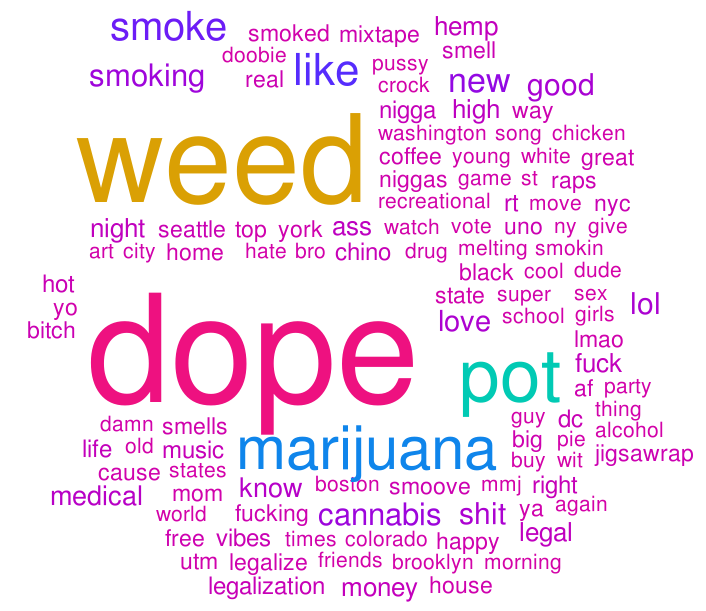}} 
\subfloat[]{\includegraphics[width=1.1\columnwidth]{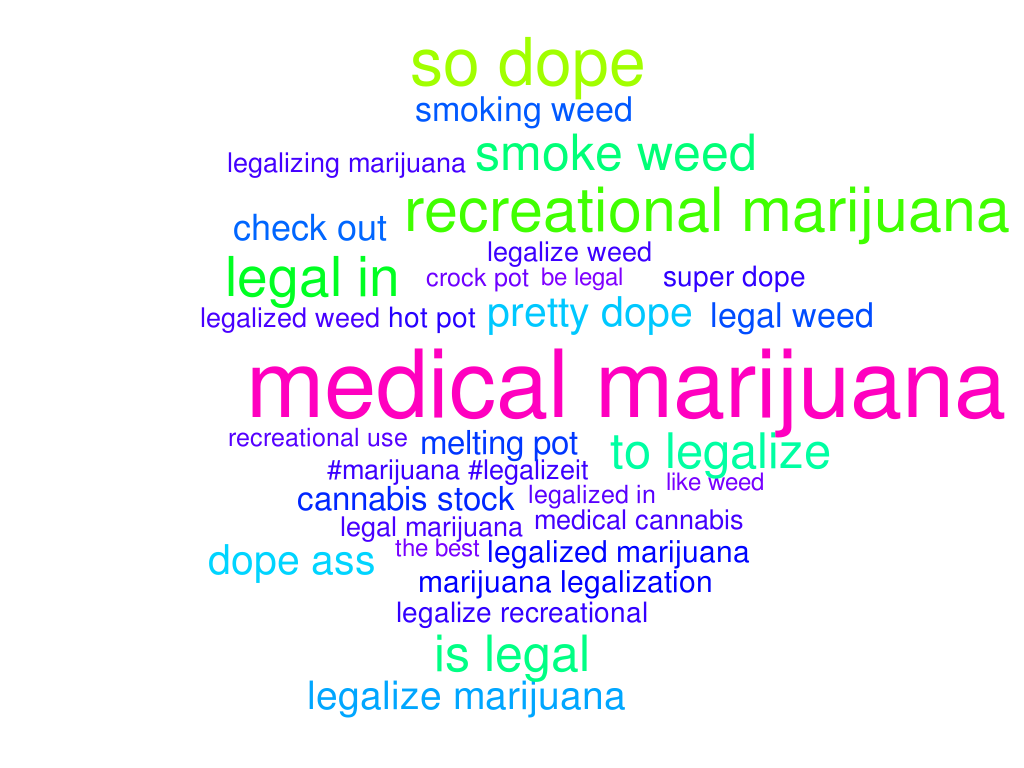} } \\

\caption{Unigram cloud (a) and Bigram cloud (b) of marijuana-related tweets collected during November 2016.}
\label{fig:mj_cloud_chart} 
\end{figure*}

%
%
 \begin{table}[t]
\centering
\caption{Top 20 users who post marijuana-related tweets with external links}
\label{tab:outerlink}
\begin{tabular}{|l|l|l|ll}
\cline{1-3}
\textbf{User}              & \textbf{Number of links} & \textbf{Place}                &  &  \\ \cline{1-3}
Potnetworkcom     & 1638           & Denver, CO           &  &  \\ \cline{1-3}
eatin\_n\_streets & 1582           & Denver, CO           &  &  \\ \cline{1-3}
DenverCP          & 764            & Denver, CO           &  &  \\ \cline{1-3}
\_DiegoPellicer\_ & 654            & Seattle, WA          &  &  \\ \cline{1-3}
MME\_MESA         & 424            & Mesa, AZ             &  &  \\ \cline{1-3}
ermphd            & 410            & Austin, TX           &  &  \\ \cline{1-3}
OG\_Chino         & 397            & Los Angeles, CA      &  &  \\ \cline{1-3}
ABG\_Marketplace  & 343            & Kansas City, MO      &  &  \\ \cline{1-3}
WeedFeed          & 271            & Chicago, IL          &  &  \\ \cline{1-3}
Boston\_CP        & 270            & Boston, MA           &  &  \\ \cline{1-3}
SLM420LOVE        & 269            & California, CA       &  &  \\ \cline{1-3}
CoCannabisCo      & 266            & Oregon, OR           &  &  \\ \cline{1-3}
SpeakEasy\_SEVL   & 252            & Colorado Springs, CO &  &  \\ \cline{1-3}
Chance\_Takers    & 243            & Atlanta, GA          &  &  \\ \cline{1-3}
greco\_james      & 238            & Phoenix, AZ          &  &  \\ \cline{1-3}
Diabetes\_Newzz   & 236            & New York, NY         &  &  \\ \cline{1-3}
PhoenixCP         & 233            & Phoenix, AZ          &  &  \\ \cline{1-3}
420digitalweb     & 224            & Denver, CO           &  &  \\ \cline{1-3}
Cannabis\_Card    & 212            & San Diego, LA        &  &  \\ \cline{1-3}
StartupCannabis   & 206            & New York, NY         &  &  \\ \cline{1-3}
\end{tabular}
\end{table}

\begin{figure}
\centering 
\includegraphics[width=1.0\columnwidth]{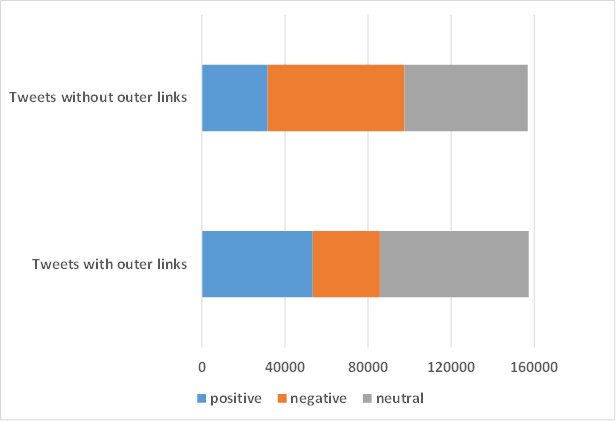}
\caption{Sentiment analysis of $158,814$ tweets without outer links and $157,377$ tweets with outer links: Tweets without outer links seem to be more negative than ones with outer links. }
\label{fig:mmj_sentiment} 
\vspace{-0.1in}
\end{figure}
\section{Results and Discussion}

\subsection{Marijuana Unigram and Bigram Clouds}
Unigrams and bigrams allow generating cloud tags for illustration of popular terms. Fig. \ref{fig:mj_cloud_chart}(a) shows the most frequent terms among unigrams (after removing some most common terms in English). Generally, ``dope",``weed", ``pot", and ``marijuana" are some highlight words. Besides those four favorite words, there are many action words associated with marijuana consumption such as ``smoke", ``smoking", ``buy", ``like", ``love" and ``smell". 
Interestingly, data extracted from our text-mining algorithm indicated that there were many terms with provocative meaning such as ``ass", ``bitch", ``shit",  or ``dam" are usually used in marijuana tweets. In addition, our data also shows strong correlation between number of tweets and some geographical locations which appeared have substantial activities related to cannabis. For example, Colorado, which has firstly legalized using marijuana for adult 21 years of age or older, is mentioned most. This suggests that legalization of marijuana in many areas has sparked a controversy, including positive and negative opinions.

\begin{figure*}
  \includegraphics[width=7in,height=3.2in]{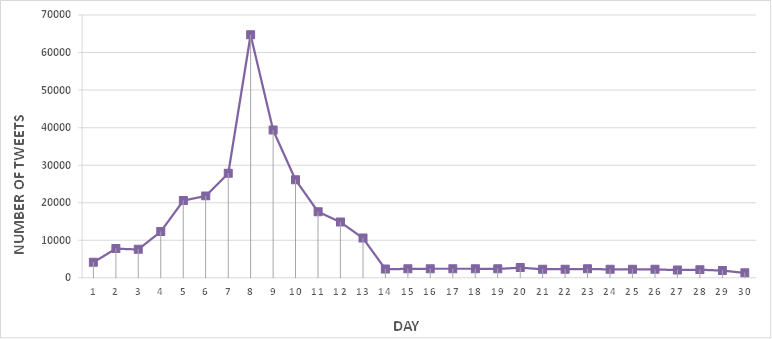}
  \caption{Daily distribution of the number of tweets relating to marijuana in November 2016: there is an exponential increase in the number of marijuana-related tweets during the week of the US presidential election and legalization votes in some more states.}
  \label{fig:mmj_tweet_on_each_day_in_month}
\end{figure*}

Fig. \ref{fig:mj_cloud_chart}(b) reveals more detailed information of marijuana use via the bigram cloud. Apparently, this type of cloud clearly shows many marijuana-related vocabularies such as legal, melting, dope, super, crock, etc. Also, the frequency of pair words which is related to legalization is very popular. It may reflect the event of state legalization votes during November.  The frequency of ``medical marijuana" indicates that more and more users want to promote the benefit of using marijuana for medical purposes.

%
%

\subsection{Identifying Users' Attitudes via Tweets}
Our data indicates that we can actually can distinguish users' attitudes towards cannabis use via the number of outer links in their tweets. Outer links or external links are identified, based on the total number of URLs in the tweet metadata including full URLs and shortened URLs. Particularly, more than 300,000 tweets in our database, there are total $158,814$ tweets without outer links and $157,377$ tweets with outer links. Table \ref{tab:outerlink} shows top 20 users who had outer links in their tweets. Our data analysis reveals that most of these users (17/20) were from states where the use of marijuana for medical purpose or recreational use is legal (e.g., Colorado, Washington, Illinois, Massachusetts, California, and New York). For example, top three users with most tweets containing external links, unsurprisingly, come from Denver, Colorado - one of the first state where marijuana is legal for both medical and recreational purpose.  More specifically, we find out that most of these users are likely to be news and magazine organizations such as Potnetworkcom, DenverCP, PhoenixCP, Boston\_CP, WeedFeed, MME\_MESA. For example, user Potnetworkcom with 1638 tweets, has a website \textit{http://potnetwork.com} - that publishes all things Marijuana and entertain other users ``with up to date information about marijuana pop culture" or DenverPC belongs to website \textit{http://toplocalnow.com/} that tweets breaking news and weather updates from Denver and many other cities. Our data reveals that many organizations, which provide services and products associated with marijuana, tend to utilize Twitter to promote their products and generate publicity.

We use a tweet sentiment analysis tool by Mashape \cite{mashape_url} to estimate the Twitter user attitude towards Cannabis. The tool works by examining individual words and short sequences of words (n-grams) and comparing them with a probability model. We analyze two sets: the tweets with outer links and tweets without outer links for evaluation. In Fig. \ref{fig:mmj_sentiment}, we present information about the proportion of the sentiment. Overall, for the set of tweets with URLs, the percentage of positive tweets is higher than the negative tweets. For the set of tweets without URLs, however, the percentage of positive tweets is much lower. Considering the group of positive tweets, the proportion of the positive tweet with external links is 62\%, compared with 32\% of tweets without external links. This implies that many users who attach external links to some websites try to deliver the information about the benefits of marijuana, such as for medical and experiments. They might want other people to perceive the advantages of marijuana. Users, who do not attach URLs in their tweets, can be individual marijuana smokers. However, because of many offensive terms (e.g., ``bitch", or alike words) included in their tweets, they are identified as having negative sentiments towards marijuana.

\begin{figure*}
\centering 
\subfloat[]{\includegraphics[width=1.0\columnwidth]{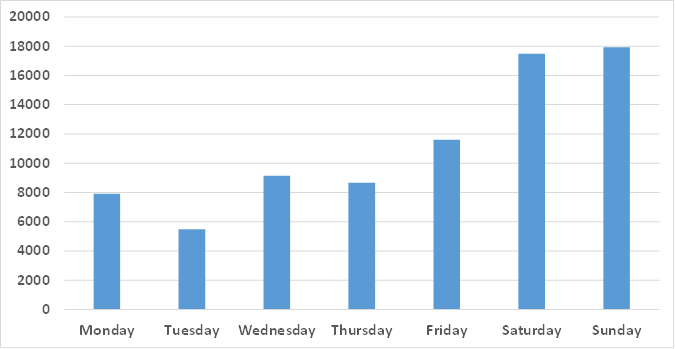}} 
\subfloat[]{\includegraphics[width=1.1\columnwidth]{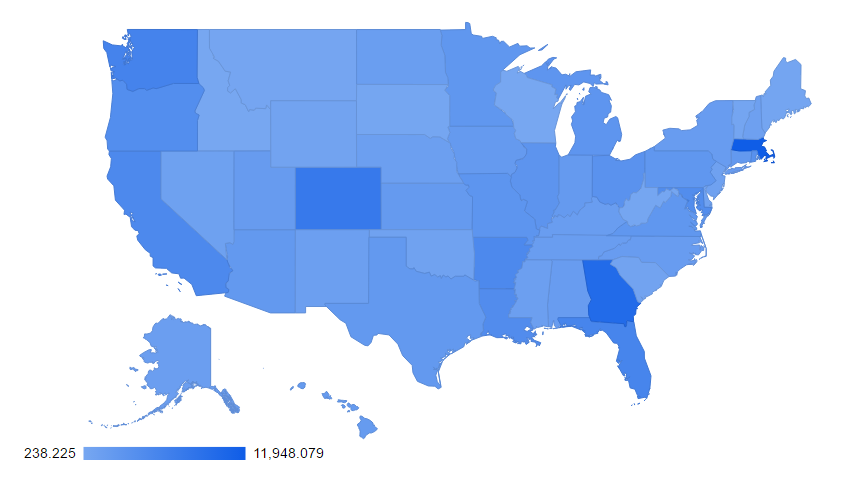} } \\

\caption{(a) Daily distribution of marijuana-related tweets in a week during November 2016 - the average for the whole month excluding the day of presidential election and marijuana legalization vote; (b) the state map of the marijuana-related tweet frequencies (the number of tweets over the population of each state). }
\label{fig:time_space_chart} 
\end{figure*}

\subsection{Temporal and Spatial Distribution of Tweets}
The volume of marijuana-related discussions is largely driven by political events. Fig. \ref{fig:mmj_tweet_on_each_day_in_month} shows the daily distribution of the number of tweets relating to marijuana in November 2016. Clearly, in the first week of the month, the number of tweets increased at an exponential rate and reached a peak on November 8. The tweets express the user's emotion and opinion about the marijuana policy reforms. For example, on November 9 four more states (California, Nevada, Maine, and Massachusetts in addition to Colorado, Washington, Alaska, Oregon, and Washington DC) voted for legal marijuana consumption of both recreational and medical purpose \cite{four_state_mmj_url}. Another important reason is that  the same period, the outcome of US presidential election was decided and the elected president has shown support for using cannabis for medical purpose and is likely to encourage the federal government to allow more states to vote on legalizing recreational marijuana \cite{hillary_trump_url}.  

It is interesting to estimate the tweet frequency during the regular weeks, i.e., without special effects of presidential elections or marijuana legalization events. We, therefore, consider the time from November 15 to 31. The research of \cite{kypri2014effects} proves there are more tweets about alcohol during the weekend. This is also true for marijuana. Fig. \ref{fig:time_space_chart}(a) presents a daily distribution of marijuana-related tweets in regular weeks, i.e., excluding the abnormal weeks of U.S. Election Day 2016 and cannabis legalization election days. We observe a clear trend that there is a significant uptick in the number of tweets at the weekend as compared to weekdays. We make a prediction that at the weekend, users tend to have more spare time to enjoy recreational activities. Also, Twitter accounts of celebrities, the media or businesses might exploit the value of weekend tweeting to post more tweets since their audiences have more time to consume and share content.

By the end of November 2016, there are eight states including California, Nevada, Maine, and Massachusetts, Colorado, Washington, Alaska, and Oregon, and Washington DC which have been legalized to use marijuana for both recreational and medical purpose \cite{mmj_governing_url}. Our spatial graph in Fig. \ref{fig:time_space_chart}(b) shows that there are more tweets from those eight states, thus matching with the marijuana state law map in \cite{mmj_governing_url}. The number of tweets, however, are also quite high in some states such as Georgia. Based on the federal laws and state marijuana laws map, we expect that fewer marijuana related tweets in this state because this area only allows for limited medical purposes. Surprisingly, our data indicated a contrary observation. This can be interpreted as there are some level of marijuana use beyond the medical purposes. Further study on such issues is considered as our future work.
\begin{figure}
\centering 
\includegraphics[width=1.0\linewidth]{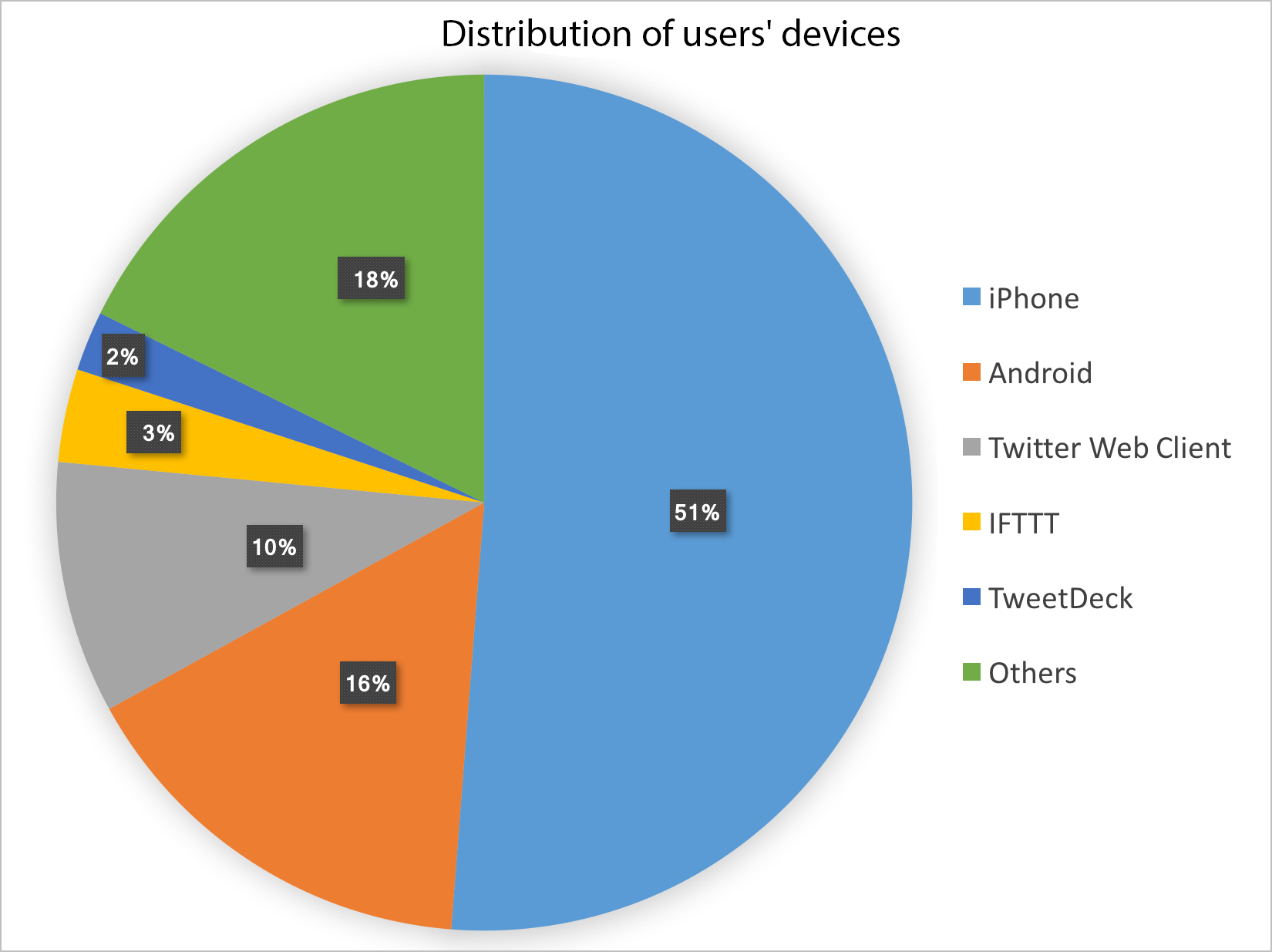} 
\caption{The proportion of devices of the users who post marijuana-related tweets on Twitter }
\label{fig:distribution_user_device} 
\vspace{-0.1in}
\end{figure}

\subsection{ Types of Devices Used for Marijuana Tweeting}

\begin{figure*}
   \includegraphics[width=7in,height=3.8in]{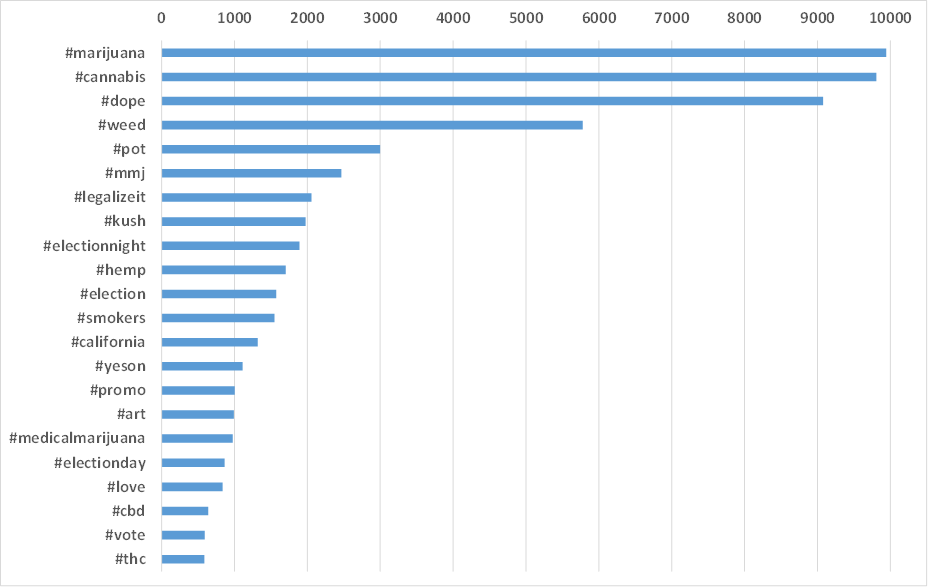}
  \caption{Top 20 hashtags in marijuana-related tweets:  \#marijuana, \#cannabis, \#dope, and \#weed are most common hashtags.}
  \label{fig:mmj_hashtag_chart}
\end{figure*}

It is known that 82\% of active users are on mobile phones \cite{twitter_url}. This raises us a question on the device types of marijuana tweeting users. Within more than 91,000 users we process, about 67\% use mobile phones (51\% for iPhone and 16\% for Android phones) via the Twitter mobile application to post their tweets (Fig. \ref{fig:distribution_user_device}). There are about $8,695$ users who use Internet browsers such as Chrome, Firefox, and Safari that are in the group of Twitter Web Client. Importantly, many users (remaining $23.5\%$) employ third-party services to publish their marijuana-related tweets. Two such popular services are IFTTT and TweetDeck. IFTTT, an abbreviation of ``If This Then That", is a web-based service that allows users to tweet automatically based on schedules or some particular events. Similarly, TweetDeck is a Twitter tool for real-time tracking, organizing, and engagement that helps users to reach their audiences by automatic postings. Thus, there is an unusually less number of users using mobile devices when comparing to the average number of 82\%. Our observed data can be explained that in the marijuana related tweets, many users are employing automated posting services to promote their products or implementing marketing strategies.

\subsection{Marijuana-related Hashtags}

Topics in Tweeter are categorized based on hashtags, labeling words or phrases preceded by pound sign (\#). By using hashtags, Twitter's users can express their tweet's content, and thus, specific subjects of discussions among users can be found more quickly. Fig. \ref{fig:mmj_hashtag_chart} illustrates most common hashtags among marijuana-related discussions. Unsurprisingly, \#marijuana \#cannabis \#dope and \#weed are the most ubiquitous terms. Besides, other marijuana-related hashtags are also frequently used such as \#pot, \#kush, \#mmj, \#hemp, \#cbd and \#thc. Two terms \#cbd and \#thc respectively refer to Cannabidiol and Tetrahydrocannabinol, two main ingredients in the marijuana plant. \#mmj means ``marijuana".

We also notice that some political hashtags frequently appear such as \#legalizeit, \#electionnight, \#electionnday, \#vote, \#YESon. \#YESon means ``Yes on", a typical phrase used by Twitter users in the election campaigns. There might be a variety of reasons for this circumstance. Firstly, our data set is collected during November 2016. During this time, nine states were voting for marijuana legalization, including Florida, Massachusetts, North Dakota, Maine, Arkansas, Montana, Arizona, Nevada, California. On November 8, California, Nevada, Maine, and Massachusetts all voted in favor of legalized use, sale, and consumption of recreational marijuana. Secondly, the appearance of political hashtags \#electionnight, \#electionnday, and \#vote reflects US presidential election event happening at the same time. It is clear that the presidential candidates' attitudes and the new government's policy toward the state marijuana legalization trend will dramatically effect every marijuana business and individual who consumes marijuana, just provoking a lot of discussions on this topic.

\section{Conclusion}
We address the challenges of unstructured Tweets' content by implementing efficient and accurate text-mining algorithms. As a result, many interesting and valuable features of data are extracted. Firstly, by analyzing the Unigrams and Bigram of tweet's content and the distribution of marijuana-related tweets within a week and a month, we reveal that the tweet's content tends to reflect the opinion of users about the current related topics such as medical marijuana, marijuana legalization, and the US presidential election. Secondly, the data also shows the geographical distribution of marijuana use across 50 states of U.S. with some unexpected observations. In addition, our result also reveals some level of association between users' attitudes and tweets with and without external links. Finally, our study spots the differences between the way and purpose of the individual users and the organizational users. Those findings would suggest some valuable patterns which could be used as a marijuana surveillance approach for federal authorities and public health agencies in developing policy and regulations.
\bibliographystyle{IEEEtran}
\bibliography{references}

\begin{thebibliography}{10}
\providecommand{\url}[1]{#1}
\csname url@samestyle\endcsname
\providecommand{\newblock}{\relax}
\providecommand{\bibinfo}[2]{#2}
\providecommand{\BIBentrySTDinterwordspacing}{\spaceskip=0pt\relax}
\providecommand{\BIBentryALTinterwordstretchfactor}{4}
\providecommand{\BIBentryALTinterwordspacing}{\spaceskip=\fontdimen2\font plus
\BIBentryALTinterwordstretchfactor\fontdimen3\font minus
  \fontdimen4\font\relax}
\providecommand{\BIBforeignlanguage}[2]{{%
\expandafter\ifx\csname l@#1\endcsname\relax
\typeout{** WARNING: IEEEtran.bst: No hyphenation pattern has been}%
\typeout{** loaded for the language `#1'. Using the pattern for}%
\typeout{** the default language instead.}%
\else
\language=\csname l@#1\endcsname
\fi
#2}}
\providecommand{\BIBdecl}{\relax}
\BIBdecl

\bibitem{nytimes_url}
\BIBentryALTinterwordspacing
M.~Isaac and S.~Ember. For election day influence, twitter ruled social media.
  [Online]. Available:
  \url{http://www.nytimes.com/2016/11/09/technology/for-election-day-chatter-twitter-ruled-social-media.html}
\BIBentrySTDinterwordspacing

\bibitem{twitter_url}
\BIBentryALTinterwordspacing
Twitter. Twitter usage / company facts. [Online]. Available:
  \url{https://about.twitter.com/company}
\BIBentrySTDinterwordspacing

\bibitem{culotta2013lightweight}
A.~Culotta, ``Lightweight methods to estimate influenza rates and alcohol sales
  volume from twitter messages,'' \emph{Language resources and evaluation},
  vol.~47, no.~1, pp. 217--238, 2013.

\bibitem{myslin2013using}
M.~Mysl{\'\i}n, S.-H. Zhu, W.~Chapman, and M.~Conway, ``Using twitter to
  examine smoking behavior and perceptions of emerging tobacco products,''
  \emph{Journal of medical Internet research}, vol.~15, no.~8, p. e174, 2013.

\bibitem{xu2016leveraging}
S.~Xu, C.~Markson, K.~L. Costello, C.~Y. Xing, K.~Demissie, and A.~A. Llanos,
  ``Leveraging social media to promote public health knowledge: Example of
  cancer awareness via twitter,'' \emph{JMIR public health and surveillance},
  vol.~2, no.~1, 2016.

\bibitem{macleod2004psychological}
J.~Macleod, R.~Oakes, A.~Copello, I.~Crome, M.~Egger, M.~Hickman,
  T.~Oppenkowski, H.~Stokes-Lampard, and G.~D. Smith, ``Psychological and
  social sequelae of cannabis and other illicit drug use by young people: a
  systematic review of longitudinal, general population studies,'' \emph{The
  Lancet}, vol. 363, no. 9421, pp. 1579--1588, 2004.

\bibitem{fried2002current}
P.~Fried, B.~Watkinson, D.~James, and R.~Gray, ``Current and former marijuana
  use: preliminary findings of a longitudinal study of effects on iq in young
  adults,'' \emph{Canadian Medical Association Journal}, vol. 166, no.~7, pp.
  887--891, 2002.

\bibitem{url_mmj_wh}
\BIBentryALTinterwordspacing
W.~House. The public health consequences of marijuana legalization. [Online].
  Available: \url{https://www.whitehouse.gov/ondcp/marijuana}
\BIBentrySTDinterwordspacing

\bibitem{pak2010twitter}
A.~Pak and P.~Paroubek, ``Twitter as a corpus for sentiment analysis and
  opinion mining.'' in \emph{LREc}, vol.~10, 2010, pp. 1320--1326.

\bibitem{agarwal2011sentiment}
A.~Agarwal, B.~Xie, I.~Vovsha, O.~Rambow, and R.~Passonneau, ``Sentiment
  analysis of twitter data,'' in \emph{Proceedings of the workshop on languages
  in social media}.\hskip 1em plus 0.5em minus 0.4em\relax Association for
  Computational Linguistics, 2011, pp. 30--38.

\bibitem{wang2014hurricane}
H.~Wang, E.~Hovy, and M.~Dredze, ``The hurricane sandy twitter corpus,''
  \emph{Links}, vol. 499, pp. 539--515, 2014.

\bibitem{kouloumpis2011twitter}
E.~Kouloumpis, T.~Wilson, and J.~D. Moore, ``Twitter sentiment analysis: The
  good the bad and the omg!'' \emph{Icwsm}, vol.~11, pp. 538--541, 2011.

\bibitem{li2016discovering}
Q.~Li, S.~Shah, R.~Fang, A.~Nourbakhsh, and X.~Liu, ``Discovering relevant
  hashtags for health concepts: A case study of twitter,'' in \emph{Workshops
  at the Thirtieth AAAI Conference on Artificial Intelligence}, 2016.

\bibitem{denecke2013exploit}
K.~Denecke, M.~Krieck, L.~Otrusina, P.~Smrz, P.~Dolog, W.~Nejdl, E.~Velasco
  \emph{et~al.}, ``How to exploit twitter for public health monitoring,''
  \emph{Methods Inf Med}, vol.~52, no.~4, pp. 326--39, 2013.

\bibitem{paul2012model}
M.~J. Paul and M.~Dredze, ``A model for mining public health topics from
  twitter,'' \emph{Health}, vol.~11, pp. 16--6, 2012.

\bibitem{cavazos2015twitter}
P.~A. Cavazos-Rehg, M.~Krauss, S.~L. Fisher, P.~Salyer, R.~A. Grucza, and L.~J.
  Bierut, ``Twitter chatter about marijuana,'' \emph{Journal of Adolescent
  Health}, vol.~56, no.~2, pp. 139--145, 2015.

\bibitem{cavazos2014characterizing}
P.~Cavazos-Rehg, M.~Krauss, R.~Grucza, and L.~Bierut, ``Characterizing the
  followers and tweets of a marijuana-focused twitter handle,'' \emph{Journal
  of medical Internet research}, vol.~16, no.~6, 2014.

\bibitem{mashape_url}
\BIBentryALTinterwordspacing
Mashape. Tweet sentiment api. [Online]. Available:
  \url{https://market.mashape.com/vivekn/sentiment-3}
\BIBentrySTDinterwordspacing

\bibitem{four_state_mmj_url}
\BIBentryALTinterwordspacing
B.~Gilbert. 4 states just voted to make marijuana completely legal — here's
  what we know. [Online]. Available:
  \url{http://www.businessinsider.com/marijuana-states-legalized-weed-2016-11}
\BIBentrySTDinterwordspacing

\bibitem{hillary_trump_url}
\BIBentryALTinterwordspacing
S.~G. .~B. Insider. Here's where donald trump and hillary clinton stand on
  marijuana legalization. [Online]. Available:
  \url{http://finance.yahoo.com/news/heres-where-donald-trump-hillary-141200233.html}
\BIBentrySTDinterwordspacing

\bibitem{kypri2014effects}
K.~Kypri, G.~Davie, P.~McElduff, J.~Connor, and J.~Langley, ``Effects of
  lowering the minimum alcohol purchasing age on weekend assaults resulting in
  hospitalization in new zealand,'' \emph{American journal of public health},
  vol. 104, no.~8, pp. 1396--1401, 2014.

\bibitem{mmj_governing_url}
\BIBentryALTinterwordspacing
G.-T. States and Localities. State marijuana laws in 2016 map. [Online].
  Available:
  \url{http://www.governing.com/gov-data/state-marijuana-laws-map-medical-recreational.html}
\BIBentrySTDinterwordspacing

\end{thebibliography}

\end{document}